*The parabolic trigonometric functions and the Chebyshev radicals*

*G. Dattoli, M. Migliorati [1] and P. E. Ricci[2]*

ENEA, Tecnologie Fisiche e Nuovi Materiali, Centro Ricerche Frascati
C.P. 65 - 00044 Frascati, Rome (Italy)

**ABSTRACT**

The parabolic trigonometric functions have recently been introduced as an intermediate step between circular and hyperbolic functions. They have been shown to be expressible in terms of irrational functions, linked to the solution of third degree algebraic equations. We show the link of the parabolic trigonometric functions with the Chebyshev radicals and also prove that further generalized forms of trigonometric functions, providing the natural solutions of the quintic algebraic equation, can be defined. We also discuss the link of this family of functions with the modular elliptic functions.

[1]  Dipartimento di Energetica , Università di Roma "La Sapienza", Via A. Scarpa, 14 - 00161 Rome, Italy

[2]  Dipartimento di Matematica, Università di Roma "La Sapienza", P.le Aldo Moro, 5 – 00185 Rome, Italy

## 1.   Introduction

The parabolic sine and cosine functions (p t f) are geometrically defined in Fig. 1 and have been shown to satisfy the identity [1]

$$\left[\cos p(\Phi)\right]^2 + \sin p(\Phi) = 1 \tag{1}$$

which will be referred to as the p t f fundamental identity.

The geometrical meaning of the argument $\Phi$ is defined in Fig. 1 and the following properties under derivative proved in [1]

$$\begin{aligned}
\frac{d}{d\Phi}\cos p(\Phi) &= -\frac{1}{\sin p(\Phi) + 2\left[\cos p(\Phi)\right]^2} = -\frac{1}{1 + \left[\cos p(\Phi)\right]^2}, \\
\frac{d}{d\Phi}\sin p(\Phi) &= 2\,\frac{\cos p(\Phi)}{\sin p(\Phi) + 2\left[\cos p(\Phi)\right]^2} = 2\,\frac{\cos p(\Phi)}{1 + \left[\cos p(\Phi)\right]^2}
\end{aligned} \tag{2}$$

can be combined to get a further identity, which can be viewed as the implicit definition of the parabolic cosine. Setting indeed $\cos p(\Phi) = Y$ we find (see ref. [1] for further details)

$$Y^3 + 3Y + 3\Phi - 4 = 0 \tag{3}.$$

The parabolic cosine is therefore described by means of the real solution $y_1$ of the above trinomial equation, or through the real part of the relevant complex solutions.

This aspect of the problem, along with the irrational (and not transcendent) nature of the p t f has been discussed in Ref. [1], here we will analyze the link between the p t f and the Chebyshev radicals [2]. We set therefore

$$\begin{aligned}
\xi &= -iY, \\
\tau &= 4 - 3\Phi
\end{aligned} \tag{4}$$

thus finding from eq. (3)

$$\xi^3 - 3\xi = i\,\tau \tag{5}$$

whose complex solution can be written in terms of Chebishev radicals $C_{1/3}(x)$ as

$$\xi = C_{1/3}[i\,\tau] = 2\cos\left[\frac{1}{3}\cos^{-1}\left(\frac{i\,\tau}{2}\right)\right] \tag{6a}$$

the parabolic cosine can therefore be written as

$$\cos p(\Phi) = -2\,\mathrm{Re}(C_{1/3}(i\tau)) \tag{6b}.$$

In the forthcoming part of the paper we will describe how the p t f can be associated, through the Chebyshev polynomials, with the hypergeometric function and how higher p t f can be used as natural solutions of algebraic equations with degree larger than the fourth.

## 2. The Chebyshev radicals and the parabolic trigonometric functions

It is to be noted that $\xi$ given by eqs. (6a) is complex, the parabolic cosine is a real quantity, therefore going back to the solution of the third order algebraic equation we use the identity [2]

$$\cos p(\Phi) = S_{1/3}[\tau] = -2\,\mathrm{Re}\left[i\,C_{1/3}(i\tau)\right] = 2\,i\sin\left[\frac{1}{3}\sin^{-1}\left(i\,\frac{3\Phi - 4}{2}\right)\right] =$$
$$= -2\sinh\left[\frac{1}{3}\sinh^{-1}\left(\frac{3\Phi - 4}{2}\right)\right] \tag{7}.$$

The Chebyshev polynomials can be shown to be the solution of a second order differential equation with non constant coefficient. The functions [3]

$$f_1(x) = 2\sin\left(\lambda\sin^{-1}\left(\frac{x}{2}\right)\right),$$
$$f_2(x) = 2\cos\left(\lambda\sin^{-1}\left(\frac{x}{2}\right)\right) \tag{8}$$

are indeed solutions of the equation

$$\left[(4 - x^2)\left(\frac{d}{dx}\right)^2 - x\frac{d}{dx} + \lambda^2\right]f(x) = 0 \tag{9}$$

for any real value of $\lambda$.

The variable transformation

$$x = 2(1 - 2\eta) \tag{10}$$

allows to recast eq. (9) as

$$\left[ \eta(1-\eta)\left(\frac{d}{d\eta}\right)^2 + \frac{1-2\eta}{2}\frac{d}{d\eta} + \lambda^2 \right] g(\eta) = 0 \tag{11},$$

which is recognized as a hypergeometric type equation, so that the solution of eq. (9) can also be written as

$$f(x) = {}_2F_1\left(\lambda, -\lambda, \frac{1}{2}; \frac{2-x}{4}\right) \tag{12}.$$

Since $C_{1/3}(i\tau)$ can be written as a linear combination of eqs. (8) with $\lambda = 1/3$ it is evident that it can be written in terms of the hypergeometric function (12) thus finding

$$C_{1/3}(\tau) = 2\,{}_2F_1\left(\frac{1}{3}, -\frac{1}{3}, \frac{1}{2}; \frac{2-\tau}{4}\right) = \sum_{n=0}^{\infty} \frac{2}{1-3n}\binom{3n}{n}\left(\frac{2-\tau}{27}\right)^n \tag{13}.$$

The parabolic cosine therefore reads

$$\cos p(\Phi) = 4\,\mathrm{Re}\left[ i\,{}_2F_1\left(\frac{1}{3}i, -\frac{1}{3}i, \frac{1}{2}; i\frac{3}{4}(2-\Phi)\right)\right] =$$
$$= 4\,\mathrm{Re}\left[ \sum_{n=0}^{\infty} \frac{i}{1-3n}\binom{3n}{n}\left(\frac{2-i(3\Phi-4)}{27}\right)^n \right] \tag{14}$$

According to the previous results we can conclude that

$$\cos p(\Phi) = -2\sinh\left(\frac{1}{3}\sinh^{-1}(\frac{3\Phi-4}{2})\right),$$
$$\sin p(\Phi) = 1 - 4\left[\sinh\left(\frac{1}{3}\sinh^{-1}\left(\frac{3\Phi-4}{2}\right)\right)\right]^2 =$$
$$= 3 - 2\cosh\left(\frac{2}{3}\sinh^{-1}\left(\frac{3\Phi-4}{2}\right)\right) \tag{15},$$

These last expressions can be exploited to gain further insight into the theory of p t f.

Before closing this section, let us note that the algebraic equation

$$x^3 + a\,x^2 + b\,x + c = 0 \tag{16}$$

can be reduced to the form (3) after setting

$$
\begin{aligned}
x &= \sqrt{p}\; Y - \frac{a}{3} \\
p &= \frac{1}{9}\left[3b - a^2\right] \\
3\Phi - 4 &= \frac{27c + 2a^3 - 9ab}{\sqrt{\left(3b - a^2\right)^3}}
\end{aligned}
\tag{17}
$$

thus finding that one of the roots of eq. (17) can be written in terms of the t p f according to the identity

$$x_1 = \frac{1}{3}\sqrt{3b - a^2}\,\cos p\left(\frac{1}{3}\left(\frac{27c + 2a^3 - 9ab}{\sqrt{\left(3b - a^2\right)^3}} + 4\right)\right) - \frac{a}{3} \tag{18}$$

The remaining roots can be expressed as function of $x_1$, so that all the solutions of eq. (16) are given in terms of p t f, as explicitly shown in the concluding section.

## 3. The generalised trigonometric functions and the solution of higher order algebraic equation

The generalised trigonometric functions are defined through the following extension of eq. (1), with $p,q$ two distinct relatively prime integers

$$
\begin{aligned}
&C(\Phi \mid p,q)^p + S(\Phi \mid p,q)^q = 1, \\
&\frac{1}{2} C(\Phi \mid p,q) S(\Phi \mid p,q) + \int_{C(\Phi \mid p,q)}^{1} \left[1 - \xi^p\right]^{\frac{1}{q}} d\xi = \frac{1}{2}\Phi
\end{aligned}
\tag{19}
$$

which, once combined, yield the derivation rules

$$\frac{d}{d\Phi}C(\Phi \mid p,q) = -q\,\frac{S(\Phi \mid p,q)^{q-1}}{qS(\Phi \mid p,q)^q + pC(\Phi \mid p,q)^p},$$
$$\frac{d}{d\Phi}S(\Phi \mid p,q) = p\,\frac{C(\Phi \mid p,q)^{p-1}}{qS(\Phi \mid p,q)^q + pC(\Phi \mid p,q)^p}$$

(20).

It is also proved by direct check that

$$\frac{d}{d\Phi}T(\Phi \mid p,q) = \frac{1}{C(\Phi \mid p,q)^2},$$
$$T(\Phi \mid p,q) = \frac{S(\Phi \mid p,q)}{C(\Phi \mid p,q)}$$

(21).

In the following we will specialize the previous general results to the particular values $p,= 4, q = 1$, to stress that the above result my open a new research line in the field of elliptic functions.

Here we want to emphasize that the functions

$$C(\Phi \mid 4,1) = \cos m(\Phi),$$
$$S(\Phi \mid 4,1) = \sin m(\Phi),$$
$$\cos m(0) = 1, \sin m(0) = 0$$

(22),

can be defined through the solution algebraic quintic equation

$$3\Psi^5 + 5\Psi + \vartheta = 0,$$
$$\vartheta = 5\Phi - 8$$

(23).

Since

$$\Psi = \cos m(\Phi)$$

(24)

If we set in eq. (23)

$$\Psi = \sqrt[4]{\frac{-5}{3}}\, \Omega,$$

$$t = \left(-\frac{3}{5}\right)^{5/4} \frac{\vartheta}{3} \tag{25}$$

we can recast it in the form

$$\Omega^5 - \Omega + t = 0 \tag{26}$$

which can be solved in terms of hypergeometric functions as follows [4]

$$\Omega = t\, {}_4F_3\left(\frac{1}{5},\frac{2}{5},\frac{3}{5},\frac{4}{5};\frac{1}{2},\frac{3}{4},\frac{5}{4};\frac{3125}{256}\,t^4\right) \tag{27}$$

thus getting

$$\cos m(\Phi) = -\frac{1}{5}(5\Phi - 8)\, {}_4F_3\left(\frac{1}{5},\frac{2}{5},\frac{3}{5},\frac{4}{5};\frac{1}{2},\frac{3}{4},\frac{5}{4};-\frac{3}{2^8}(5\Phi - 8)^4\right) \tag{28}.$$

Equation (28) provides the link between the solution of a quintic equation, in the Bring-Jerrard form, and the $C(\Phi;5,1)$ trigonometric function.

It is also evident that one of the solution of the algebraic equation

$$x^5 + px + \lambda = 0 \tag{29}$$

can be written as

$$x = \sqrt[4]{\frac{3p}{5}}\, \cos m\left(\frac{54\sqrt[4]{\frac{5}{3p}}\,\lambda + 8}{5}\right), \quad p > 0 \tag{30}.$$

Further comments will be presented in the forthcoming section.

## 4. Concluding remarks

In sec. 2 we have seen how one of the solution of a third degree algebraic equation can be expressed through the parabolic cosine. Here we will discuss how the other solutions of e.g. eq. (17) be expressed in terms of $\cos p(\Phi)$. Assuming for simplicity $a = 0$ and using the relations among the roots and the coefficients of the algebraic equation, we find

$$
\begin{aligned}
&x_1 + x_2 + x_3 = 0, \\
&x_1 x_2 x_3 = -c
\end{aligned}
\qquad (31).
$$

Assuming that $x_1$ is known and given by eq. (19), solving for $x_{2,3}$ we get

$$
\begin{aligned}
x_1 &= \sqrt{\frac{b}{3}} \cos p(\beta(b)), \\
x_{2,3} &= \frac{1}{2}\sqrt{\frac{3}{b}} \sec p(\beta(b)) \left[ \frac{b}{3}\left[\cos p(\beta(b))\right]^2 \pm 2c \sqrt{\left(\frac{\cos p(\beta(b))}{\sqrt{2c}}\right)^4 - 1} \right], \\
\beta(b) &= \frac{1}{3}\left( \frac{27c}{\sqrt{(3b)^3}} + 4 \right)
\end{aligned}
\qquad (32).
$$

Analogous results, albeit in a more complicated form, can be obtained for the quintic.

It is, therefore, worth mentioning the fact that the point of view we have developed allows to get an important link between the $C(\Phi \mid 4,1)$ trigonometric function and the modular elliptic functions.

The solution of the quintic equation in the Bring form (26) can be solved using the modular elliptic functions. The method traces back to Hermite [5], who showed that the solution of eq. (26) depend on a rather complicated combination of the function

$$
\phi(\tau) = \sqrt[8]{f(\tau)}
\qquad (33)
$$

where

$$f(\tau) = \frac{\vartheta_2^4(0 \mid \tau)}{\vartheta_3^4(0 \mid \tau)} \tag{34}$$

is one of the modular elliptic functions, expressed in terms of the Jacobi $\vartheta$-functions [4].

The relevant solution is considerably complex and the comparison with the method discussed in this paper will be discussed in a separated publication. We have touched this aspect of the problem for completeness sake and to stress again the wealth of research field opened by this family of generalized trigonometric functions.

Before closing this section let us mention the further family of p t f satisfying the fundamental identity[3]

$$S^2 + C = 1 \tag{35},$$

whose relevant derivative write (see Fig. 2)

$$C' = \frac{2S}{7S^2 - 1},$$
$$S' = -\frac{1}{7S^2 - 1} \tag{36}.$$

From the second of which we get that the above p t f are implicitly defined through the cubic equation

$$\frac{7}{3}S^3 - S = \Phi \tag{37}.$$

The relevant solution, defining the "hyper-parabolic" sine writes in terms of Chebyshev radicals

---

[3] Note that
$$S = S(\Phi \mid 1,2),$$
$$C = C(\Phi \mid 1,2)$$
furthermore the link existing with the t p f is analogous to that existing between circular and hyperbolic functions (see Ref. (1))

$$S = \sqrt{\frac{3}{7}}\, C_{\frac{1}{3}}\left(-\sqrt{\frac{7}{3}}\,\Phi\right) \tag{38}$$

or in terms of nested radicals [6] as

$$S = \left[\left\{-\frac{3}{7}\Phi\right\}_n ; \left\{\frac{3}{7}\right\}_n \,\Big|\, \frac{1}{3}\right] \tag{39}$$

where

$$\left[\{a\}_n ; \{b\}_n \,\Big|\, \frac{1}{m}\right] = \sqrt[m]{a + b\sqrt[m]{a + b\sqrt[m]{a + \ldots}}} \tag{40}.$$

Equation (39) represents a further expansion of t p like functions.

In a forthcoming investigation we will discuss more deeply the expansion in terms of nested radicals.

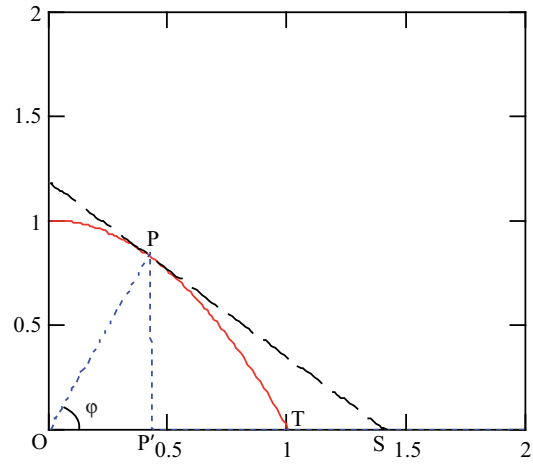

Fig. 1

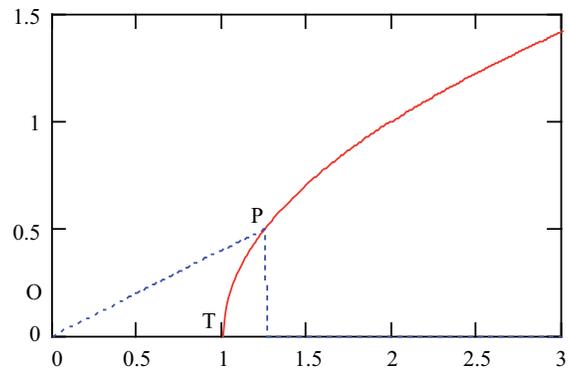

Fig. 2

**Figure captions**

Fig. 1  Geometrical interpretation of the parabolic functions. The continuous curve is an arc of hyperbola ($y+x^2=1$), the dash straight line is tangent to the hyperbola at the point P. The sector $S_{OPT}$ represents the area of the "triangle" OPT

$OP' = \cos p(\Phi), PP' = \sin p(\Phi),$

$\Phi = 2S_{OPT}$

Fig. 2  Geometrical interpretation of the parabolic functions. The continuous curve is an arc of parabola ($y^2 + x = 1$). The sector $S_{OPT}$ represents the area of the "triangle" OPT

$OP' = \cos p(\Phi), PP' = \sin p(\Phi),$

$\Phi = 2S_{OPT}$